\documentclass[aps,amsmath,amssymb,showpacs,showkeys]{revtex4-2}
\usepackage[dvips]{graphicx}
\usepackage{times}
\usepackage{braket}
\usepackage{xcolor}
\usepackage{orcidlink}
\usepackage{hyperref}
\hypersetup{
  colorlinks=true,
  urlcolor=blue,
  linkcolor=red,
  citecolor=blue
}

\begin{document}
\title{Probing $\boldsymbol{\Lambda}$CDM-mimicking $\boldsymbol{f(Q)}$ gravity 
model using gravitational waves from compact binary coalescences}

\author{Sikha Gogoi\orcidlink{0009-0000-8433-788X}}
\email[Email: ]{sikhagogoi001@gmail.com}

\author{Umananda Dev Goswami\orcidlink{0000-0003-0012-7549}}
\email[Email: ]{umananda@dibru.ac.in}

\affiliation{Department of Physics, Dibrugarh University, Dibrugarh 786004, 
Assam, India}

\begin{abstract}
The direct detection of gravitational waves (GWs) is a very significant 
achievement in the history of physics and has opened a new window to probe 
the possible deviations of physics from that of general relativity (GR). In 
this work, we forecast constraints on the free parameter of an $f(Q)$ gravity 
model that mimics a $\Lambda$CDM background at the level of cosmic expansion. 
We consider modified gravitational wave signals from inspiraling of compact 
bianry systems such as binary black holes (BBH), binary neutron stars (BNS) 
and black hole neutron star binary (BBHNS) systems in the context of the 
$f(Q)$ gravity model and perform parameter estimation for two future 
third-generation ground-based GW detectors, namely Einstein Telescope (ET) and 
Cosmic explorer (CE), respectively. Our results show that both detectors 
can give tight constraints on the model parameter up to a significantly high 
redshift. These results show the potential of future GW observations to probe 
the deviations of the nature of GWs from that of GR within the framework of 
$f(Q)$ gravity.
\end{abstract}


\maketitle                                                                      

\section{Introduction}
General Relativity (GR), which is based on the curvature of spacetime, has 
been remarkably successful in the context of its validation from various 
experimental observations and its compact mathematical structure. In $1916$, 
Einstein himself predicted the existence of gravitational waves (GWs) 
as a consequence of his theory of GR \cite{GW}. GWs are ripples in spacetime 
caused by some of the most violent and energetic processes in the Universe, 
such as the inspiral and collision of binary black holes, binary neutron stars, 
and black hole and neutron star binary systems. After almost a hundred years of
prediction, on September $14$, $2015$, GWs were finally detected by the Laser 
Interferometer Gravitational Observatory (LIGO) system, which for the first 
time detected a GW signal, named as GW$150914$, from the merger of two black 
holes \cite{GW_detection}. This event is marked as a historical achievement 
in science in general and physics in particular. Another remarkable achievement 
in support of GR is the first and subsequent imaging of the supermassive black 
holes by the Event Horizon Telescope (EHT) Collaboration at the center of M$87$ 
galaxy, as well as one at the center of our Milky Way galaxy \cite{M87}. 
However, several theoretical and observational challenges, such as issues of 
accelerated expansion of the Universe \cite{AccEx1} and galactic flat rotation 
curves \cite{missingmass, missingmass2}, cosmological constant problem 
\cite{cosmological_constant}, Hubble 
tension \cite{Hubble_tension}, etc., remain unresolved within the ambit of GR. 
Moreover, GR is a classical theory of gravity, and it can not be quantized with
the usual approach, which leads to some other issues \cite{non_renormalizable}. These issues 
have motivated the exploration of different theories of gravity. Among these 
theories, the Alternative Theories of Gravity (ATGs) belong to an interesting 
class of gravity theories, which offer different geometrical and dynamical 
descriptions of gravity \cite{ATG1,ATG2,ATG3,ATG4,fQ_1,fQ_2,fQ_3}. Within 
ATGs, the $f(Q)$ gravity, an extension of Symmetric Teleparallel Gravity 
(STG), has gained significant attention in recent years \cite{fQ_1,fQ_2,fQ_3, 
fQ_6,fQ_7,fQ_8,fQ_9,fQ_10,fQ_11,fQ_12,fQ_13,fQ_14,fQ_15,fQ_16}. In this 
framework, gravity is described entirely by the non-metricity of spacetime, 
offering a geometric interpretation different from that of curvature and 
torsion-based theories of gravity.

GWs provide a powerful and independent probe of the underlying theory of 
gravity. As mentioned, the direct detection of GWs marked a breakthrough 
in observational physics and is a very significant achievement in the history 
of physics, which opened a new window to explore the Universe in a way that 
was not possible earlier, enabling direct access to the highly dynamical and 
strong field regime of gravity. Since the first detection, numerous GW events 
have been detected by the current detectors of LIGO \cite{LIGO}, Virgo 
\cite{Virgo}, and KAGRA \cite{KAGRA}, all of which are found to be consistent 
with the prediction of GR within current observational uncertainties 
\cite{GW_BH_detection, GW_BH_detection2, GW_BH_detection3, GW_BH_detection4, 
GW_NS_detection}. However, the future third-generation ground-based GW 
detectors, such as Einstein Telescope (ET) \cite{ET1}, Cosmic Explorer (CE) 
\cite{CE1} as well as the space-based detector, Laser Interferometer Space 
Antenna (LISA) \cite{LISA} are expected to achieve significantly improved 
sensitivity across the GW frequency spectrum, which will enable us to test 
gravity with high precision and to search for possible signatures of physics 
beyond GR. 

GWs from the compact binary coalescences, such as binary black holes and binary 
neutron stars, act as ``standard sirens", enabling a direct measurement of 
luminosity distance to the source without the need for an external 
calibration \cite{ss0,ss01}. This makes them an independent tool to probe the 
expansion history of the Universe and to test ATGs as well as other theories 
of gravity on cosmological scales.

It is found in the literature that different properties of GWs, such as 
polarization mode and propagation speed, may vary in different Modified 
Theories of Gravity (MTGs) or ATGs \cite{MT1,MT2,MT3}. However, in the context 
of the modification of GR in MTGs or alternative geometrical structures of 
spacetime in ATGs, the deviation from GR not only arises in the speed and 
generation of GWs, but also in their propagation across the cosmological 
background. In particular, it is seen that the modification in the underlying 
gravity theory leads to a modified luminosity distance of GWs, which is not 
equal to the standard luminosity distance of electromagnetic radiation as in 
the case of GR \cite{luminosity_distance1, luminosity_distance2}. Such 
modifications directly affect the observed GW amplitude, as the amplitude of 
GWs scales inversely with the luminosity distance. Therefore, such effect 
directly helps in probing ATGs or MTGs.

Recent studies have explored the use of standard sirens to forecast 
constraints on $f(Q)$ gravity theories by generating mock catalogs of GW 
events, and performing the Bayesian analyses \cite{ss1, ss2,ss3, ss4}. 
However, in this work we constrain a $f(Q)$ model parameter, which mimickes
the $\Lambda$CDM model, directly at the level of gravitational waveform by 
incorporating the modification to the signal in the context of $f(Q)$ gravity. 
It is also found in the literature that using this waveform based approach, 
one can have stronger constraints even in the scenarios with limited number of 
detected events, as it fully exploits the information encoded in the waveform 
\cite{fisher1, fisher2, fisher3}. Therefore, this approach provides an 
efficient framework to investigate the deviation from GR by assuming 
modification directly in the waveform.

This paper is organized as follows: In the next Section \ref{sec2}, we give a 
brief overview of the $f(Q)$ gravity theory that is being used in this 
work. In Section \ref{sec3}, we quantify the modified GW propagation in an 
$f(Q)$ gravity model as compared to GR. In Section \ref{sec4}, we discuss 
the statistical methodology used in this work. In Section \ref{sec5}, we 
present our results of parameter estimation for GW observations with ET and CE,
respectively, and in Section \ref{sec6}, we finally provide the general summary 
and conclusion of our work.
 
\section{Basics of $\boldsymbol{f(Q)}$ gravity theory and the model}
\label{sec2}
In this section, we briefly review the formalism of $f(Q)$ gravity theory
and introduce the model used in our study. 
$f(Q)$ gravity theory is an ATG in which gravity is described by non-metricity 
of the flat torsion-free spacetime, whose action is given by \cite{fQ_2}
\begin{equation}\label{eq1}
S = \int d^4x \, \sqrt{-g}\left[ \frac{1}{2}f(Q)  + \mathcal{L}_m\right],
\end{equation}
where $\mathcal{L}_m$ is the matter Lagrangian, $g$ is the determinant of 
the metric tensor $g_{\mu\nu}$, $f(Q)$ is an arbitrary function of the 
non-metricity scalar $Q$. Here, $8{\pi}G=c=1$ is considered, and throughout 
this paper we will consider the metric signature $(-,+,+,+)$. By setting 
$f(Q)= Q$, we can get the Symmetric Teleparallel theory Equivalent of GR 
(STEGR). As mentioned previously, this theory is based on two conditions of 
spacetime, viz., symmetric teleparallelism (i.e., the curvature tensor 
$R^\alpha{}_{\beta\mu\nu}=0$) and the non-metricity condition 
($\nabla_\alpha g_{\mu\nu} \ne 0$). The non-metricity condition leads to the
non-metricity tensor as given by
\begin{equation}
\label{eq2}
Q_{\alpha\mu\nu}= \nabla_{\alpha}g_{\mu\nu}.
\end{equation}
The non-metricity scalar $Q$ is defined as the trace of non-metricity 
tensor \cite{fQ_4}, i.e.,
\begin{equation}
Q = Q_{\alpha\mu\nu}P^{\alpha\mu\nu},
\end{equation}
where $P^{\alpha}{}_{\mu\nu}$ is the non-metricity conjugate, known as the
superpotential tensor, and is given by
\begin{equation}
P^{\alpha}{}_{\mu\nu}= -\,\frac{1}{2} L^{\alpha}{}_{\mu\nu}
+ \frac{1}{4}\big( Q^{\alpha} - \tilde{Q}^{\alpha} \big) g_{\mu\nu}
- \frac{1}{4}\, \delta^{\alpha}_{(\mu} Q_{\nu)}
\end{equation}
with $L^{\alpha}{}_{\mu\nu}$ is another tensor, which is termed as the 
disformation tensor, defined by
\begin{equation}
L^{\alpha}{}_{\mu\nu}
= \frac{1}{2}\, g^{\alpha\beta}(Q_{\beta\mu\nu} - Q_{\mu\beta\nu} - 
Q_{\nu\beta\mu}),
\end{equation}
and $Q_{\alpha}=Q_{\alpha\mu\nu}g^{\mu\nu}$ and 
$\tilde{Q}_{\alpha}=\tilde{Q}_{\mu\alpha\nu}g^{\mu\nu}$ are two independent 
non-metricity vectors.

By varying the action \eqref{eq1} with respect to the metric tensor 
$g_{\mu\nu}$ under the least action principle (${\delta}S=0$) and then applying
the symmetric teleparallelism condition, the field equations of the theory can
be obtained as \cite{fQ_cosmology} 
\begin{equation}
f_Q \mathring{G}_{\mu\nu} + \frac{1}{2}\, g_{\mu\nu}\big( f_Q Q - f(Q)\big)
+ 2 f_{QQ} \mathring{\nabla}_{\alpha} Q P^{\alpha}{}_{\mu\nu} = T_{\mu\nu},
\end{equation} 
where the Einstein tensor $\mathring{G}_{\mu\nu}$ and the derivative
$\mathring{\nabla}_{\alpha}$ are for the usual Levi-Civita connection
$\mathring{\Gamma}^\alpha{}_{\mu\nu}$ as the associated affine connection in 
the theory is written as $\Gamma^\alpha{}_{\mu\nu} = 
\mathring{\Gamma}^\alpha{}_{\mu\nu} + L^{\alpha}{}_{\mu\nu}$. 
$f_Q$ and $f_{QQ}$ are respectively the first and second order derivatives
of $f(Q)$ with respect to $Q$, and $T_{\mu\nu}$ is the energy-momentum tensor 
of the matter field. Again, varying the action \eqref{eq1} with respect to the 
connection $\Gamma^\alpha{}_{\mu\nu}$, we may obtain other field equations of
$f(Q)$ gravity as
\begin{equation}
\nabla_{\mu} \nabla_{\nu}
\big( \sqrt{-g} \, f_Q \, P^{\mu\nu}{}_{\sigma}\big)= 0.
\end{equation}

In this work, we consider a homogeneous, isotropic and spatially flat FLRW 
spacetime, whose metric takes the form:
\begin{equation}\label{flrw}
ds^2 = -\,dt^2 + a(t)^2\big(dr^2 + r^2 d\theta^2 + r^2 \sin^2\theta \, d\phi^2\big),
\end{equation}
where $a(t)$ is the scale factor. We also consider that the Universe is 
composed of pressureless dust $(p=0)$. The non-metricity scalar $Q$ 
corresponding to the FLRW metric \eqref{flrw} is related to the Hubble 
parameter as \cite{fQ_5}
\begin{equation}
Q = -\,6H^2,
\end{equation}
and the modified Friedmann equations are given by \cite{fQ_cosmology}
\begin{subequations}\label{eq:friedmann}
\begin{align}
3H^2 f_Q + \frac{1}{2} \left(f(Q) - Q f_Q\right) & = \rho_m, \label{eq:friedmann_a} \\[5pt]
-\,2\, \frac{d}{dt}\big(H f_Q\big) - 3H^2 f_Q - \frac{1}{2} \big(f(Q) - Q f_Q\big) & = 0. \label{eq:friedmann_b}
\end{align}
\end{subequations}
Now, to move further, we consider a specific $f(Q)$ gravity model that gives a 
background evolution similar to that of GR, and is of the form \citep{fQ_5}:
\begin{equation}\label{model}
f(Q)=-\,2\Lambda+Q+\beta\sqrt{-Q},
\end{equation}
where $\Lambda$ is the cosmological constant and $\beta$ is a free model 
parameter that characterizes the deviation from STEGR. In the limit 
$\beta = 0$ the theory reduces to STEGR, which provides a geometric 
description of gravity based solely on the non-metricity scalar $Q$. This 
particular form of $f(Q)$ gravity theory is indistinguishable from the 
$\Lambda$CDM model at the background level, while deviations from GR may arise 
at the perturbation level. Using equations \eqref{eq:friedmann_a} and 
\eqref{eq:friedmann_b} for this particular model, we get the Hubble parameter 
similar to that of the {$\Lambda$}CDM background, as given by
\begin{equation}
H(z)= H_0 \sqrt{\Omega_{m0}(1+z)^3+(1-\Omega_{m0})},
\end{equation}
where $H_0$ is the Hubble constant and $\Omega_{m0}=\rho_{m0}/3H^2_0$ is the 
current matter density parameter.

\section{Gravitational waves propagation in $\boldsymbol{f(Q)}$ gravity}
\label{sec3}
\begin{figure}[!h]
\vspace{5pt}
    \centerline{
      \includegraphics[scale = 0.78]{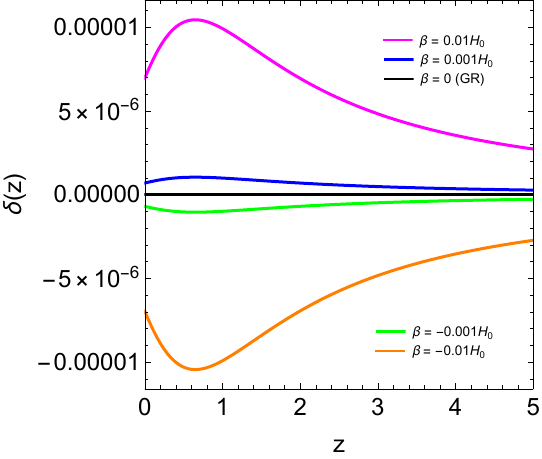}\hspace{1.0cm}
      \includegraphics[scale = 0.78]{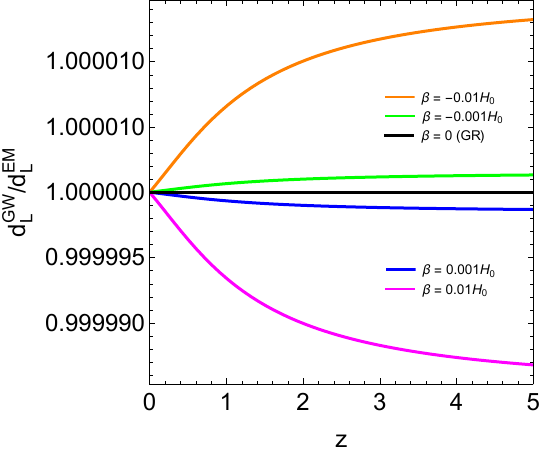}}
      \vspace{-0.2cm}
    \caption{Evolution of the characteristic quantities associated with the 
modified gravitational wave propagation as a function of redshift $z$ for 
different values of the model parameter $\beta$. Left panel shows the 
deviation $\delta(z)$ of modified friction term from GR, while the right 
panel shows the luminosity distance ratio 
$d_\text{L}^\text{\,GW}\!\!/d_\text{L}^\text{EM}$.}
\label{fig1}
\end{figure}

We now focus on the tensor perturbations around the spatially flat FLRW 
spacetime in $f(Q)$ gravity theory, which represent the GWs on cosmological 
background in the framework of the theory. In the linear perturbation theory, 
these transverse traceless tensor modes evolve independent 
of the vector and scalar perturbations, and satisfy a modified wave equation 
in any modified or alternative theory of gravity due to the nonunity value of 
the term $f_X$ in the models of such a theory, where $X$ is a general scalar 
variable of the geometry of spacetime, such as $R$ (Ricci scalar), $Q$, etc. 
Thus, in $f(Q)$ gravity theory, the evolution of the tensor perturbations is 
governed by the following modified wave equation \cite{fQ_cosmology}:
\begin{equation}
\tilde{h}_\lambda'' + 2{\mathcal{H}} \Big( 1 + \frac{d \ln \!f_Q}{2\mathcal{H} d\tau}\Big) \tilde{h}_\lambda' + k^2 \tilde{h}_\lambda = 0,
\end{equation}
where $\tilde{h}_\lambda(\tau,k)$ represents the Fourier modes of GW 
amplitude as a function of the conformal time $\tau$ and wave vector $k$ with 
$\lambda$ as the labels for the two tensor polarization modes: 
$+$ and $\times$, prime denotes derivative with respect to conformal time 
$\tau$, and $\mathcal{H}=a'/a$. In general, theories that follow the relation:
\begin{equation}
\tilde{h}_\lambda'' + 2{\mathcal{H}} \big( 1 -{\delta}(\tau)\big) \tilde{h}_\lambda' + k^2 \tilde{h}_\lambda = 0,
\end{equation}
the friction term is modified, and hence the amplitude damping of GWs differs 
from that in GR, where the function ${\delta}(\tau)$ parametrizes the amount 
of deviation from GR \cite{luminosity_distance2}. In our case,
\begin{equation}
\delta(\tau)=-\,\frac{1}{2\mathcal{H}}\frac{d\ln\! f_Q}{d\tau}.
\end{equation}
Expressing this function in terms of cosmological redshift $z$ and using the
model \eqref{model}, one can write,
\begin{equation}\label{delta}
\delta(z) 
 = \frac{1}{2}\,(1+z)\,\frac{d}{dz}\ln\!\left(1- \frac{\beta}{2\sqrt{6}H(z)}\right).
\end{equation}
The left panel of Fig.~\ref{fig1} shows the deviation $\delta(z)$ of the 
friction term as a function of redshift $z$ for different values of the model 
parameter $\beta$. From the figure, one can see that the deviation from GR is 
the model parameter $\beta$ dependent (as obvious from 
equation \eqref{delta}), which is negligible for small values of $\beta$. 
Moreover, at high redshift $z$, the deviation is very small, which increases 
with decreasing value of $z$ and reach to a maximum at intermediate redshift 
($z \sim 0.5-1$), and then decreases with decreasing $z$ towards $z<<1$, 
indicating that the effect of the $f(Q)$ gravity model \eqref{model} is more 
significant at intermediate redshifts and approaches GR at both very low 
redshifts ($z<<1$) as well as very high redshifts. Observational 
analyses have shown that the region $0.61 < z < 0.82$ is the transition 
redshift region, i.e., the redshift of transition from decelerated to 
accelerated phase of the Universe \cite{transition_redshift}. So the peak 
around the intermediate redshift may be described by the transition epoch.
 
In such a modified or alternative gravity theory, unlike GR, the luminosity 
distance for GW ($d^\text{\,GW}_\text{L}$) is not equal to that for 
electromagnetic radiation ($d^\text{\,EM}_\text{L}$), and the GW amplitude of a 
coalescing binary is proportional to $1/d^\text{\,GW}_\text{L}$ 
\cite{luminosity_distance1, luminosity_distance2}, where
\begin{equation}
\label{eq17}
d_\text{L}^\text{\,GW}(z)=d_\text{L}^\text{\,EM}\exp\!\Big[-\int_0^z\!\! \frac{\delta(z')}{1+z'} dz'\Big]
\end{equation}
with
\begin{equation}
d_\text{L}^\text{\,EM}(z) = (1+z)\! \int_0^z\!\! \frac{dz'}{H(z')}.
\end{equation}
Substituting the expression for $\delta(z)$ in equation \eqref{eq17}, the 
ratio of the luminosity distance for GW to that for EM radiation can be 
obtained as
\begin{equation}
\frac{d_\text{L}^\text{\,GW}}{d_\text{L}^\text{\,EM}}=\sqrt{\frac{f_Q(0)}{f_Q(z)}}
= \sqrt{\frac{1-\beta/2\sqrt{6}H_0}{1-\beta/2\sqrt{6}H(z)}}.
\end{equation}
In the right panel of Fig.~\ref{fig1} we plot this ratio as a function of 
redshift $z$ for different values of $\beta$. From the plot, it is evident 
that for positive (negative) values of $\beta$ and hence for $\delta >0$ 
$(<0)$, the luminosity distance of GW is less (greater) than that of EM 
radiation. Since, the amplitude of GWs is proportional to $1/d_L^\text{\,GW}$, 
therefore $\beta>0$ $(<0)$ induces decrease (increase) in GW amplitude. Also, 
one can notice that at $z<<1$, $d_L^\text{\,GW} \approx d_L^\text{\,EM}$ as
expected from the $\delta(z)$ behaviour.

\section{Fisher analysis}
\label{sec4}
In this section, we briefly discuss the analysis method employed for 
parameter estimation in this work. The bounds on the free model parameter 
${\beta}$ are obtained using GW observations for different binary coalescence 
sources. For this purpose, we employ the Fisher analysis technique described 
in \cite{fisher4, fisher5, fisher6}, which is widely used to forecast 
constraints on model parameters for future GW detectors.

In the Fisher analysis, the likelihood probability function is assumed to be 
Gaussian around its peak and the signal behaviour is approximated about that 
peak through Taylor expansion. If $\tilde{h}(\upsilon,\theta_i)$ is the GW 
form, where $\upsilon$ is the frequency and $\theta_i$ are the free 
parameters, then the root mean squared error on a parameter is
\begin{equation}
\label{eq20}
{\Delta}\theta^i=\sqrt{\Gamma_{ii}^{-1}},
\end{equation}
where $\Gamma_{ij}$ is the Fisher matrix as given by
\begin{equation}
\label{eq21}
\Gamma_{ij} = \left( \frac{\partial \tilde{h}}{\partial \theta_i} \,\middle|\, \frac{\partial \tilde{h}}{\partial \theta_j} \right).
\end{equation}
In the above expression, the notation $(a|b)$ denotes the inner product of
two vectors $a$ and $b$, defined using the detector's power spectral density 
$S_n(\upsilon)$ as
\begin{equation}
(a|b) := 2 \int_{\upsilon_\text{low}}^{\upsilon_\text{high}} 
\frac{a^*b + b^*a}{S_n(\upsilon)}\,d\upsilon,
\end{equation}
where $\upsilon_\text{high(low)}$ represents the high (low) cutoff frequency 
of the detector. From the definition of inner product, the signal-to-noise 
ratio (SNR) of the detector can be defined as
\begin{equation}
\rho^2 := (\tilde{h}|\tilde{h}) = 4 \,\mathrm{Re}\! \int_{\upsilon_\text{low}}^{\upsilon_\text{high}} \frac{|\tilde{h}(\upsilon)|^2}{S_n(\upsilon)}\,d\upsilon.
\end{equation}
In this work, we consider the TaylorF2 post-Newtonian (PN) waveform, a 
frequency-domain approximant, to describe the inspiral phase of compact 
binary coalescences with aligned spins \cite{PN_waveform} as
\begin{equation}
\label{eq24}
\tilde{h}(\upsilon) = \mathcal{A}\,\upsilon^{-7/6}\exp\bigg[i\Big(2\pi \upsilon t_c-\phi_c-\frac{\pi}{4}+\frac{3}{128\eta\nu^5}\big(1+\sum_{i\,=\,2}^{7} \alpha_i \nu^i\big)\Big)\bigg],
\end{equation}
where the amplitude $\mathcal{A}$ is given by
\begin{equation}
\mathcal{A}\propto \mathcal{M}_c^{5/6}/d_L^\text{\,GW},
\end{equation}
$\alpha_i$'s are corrections upto $3.5$ PN order that contains the spins 
$\chi_1$, $\chi_2$ of the members of binary coalescence, $t_c$ and $\phi_c$ 
are time and phase of coalescence, respectively. If $m_1$ and $m_2$ are the 
masses of the members of a binary coalescence, then $M=m_1+m_2$ is the total 
mass of the system, $\eta=m_1m_2/M^2$ is the symmetric mass ratio, 
$\mathcal{M}_c=(1+z)M\eta^{3/5}$ is the redshift chirp mass, and 
$\nu=(\pi M \upsilon)^{1/3}$ is the inspiral reduced frequency of the system. 
Therefore, the waveforms in equation \eqref{eq24} depend on the free 
parameters,
\begin{equation}
\theta_i=(M,\eta,\chi_1,\chi_2,t_c,\phi_c, \beta/H_0),
\end{equation}
where $\beta/H_0$ is the dimensionless model parameter that characterizes the 
deviation with respect to GR, as already mentioned. This parameter enters only 
through amplitude via luminosity distance $d_L^\text{\,GW}$, which means that 
it modifies the waveform in terms of amplitude only.

\section{Fisher forecast for the third generation ground-based gravitational 
wave detectors}
\label{sec5}
\begin{figure}[!h]
    \centerline{
    \includegraphics[scale = 0.55]{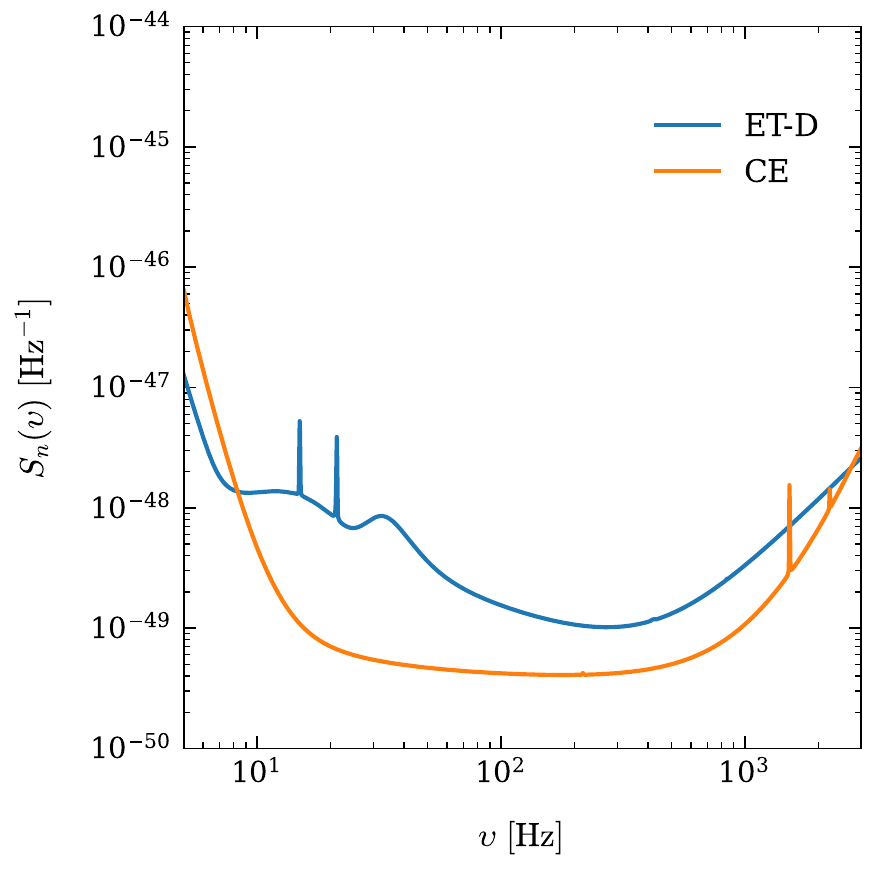}}
    \vspace{-0.2cm}
    \caption{Power Spectral Density (PSD) comparison between ET and CE 
detectors.}
    \label{fig2}
\end{figure}

In order to move for further analysis, we consider two future planned third
generation ground-based GW detectors, viz., the Einstein Telescope (ET) 
\cite{ET2, ET3, ET4} and Cosmic Explorer (CE) \cite{CE1}, with sensitivities 
shown in Fig.~\ref{fig2}. ET is a next-generation GW detector proposed in 
$2010$ \cite{ET1}. It is designed to significantly improve the sensitivity 
and frequency coverage from current detectors. The ET is expected to cover the 
frequency band in the range $1-10^4$ Hz, with an improvement in amplitude 
sensitivity of about an order of magnitude compared to current detectors. On 
the other hand, CE is also a next generation ground-based detector developed 
by American institutions, a few years after ET. CE has a network of L-shaped 
interferometers, with arm length upto $40$ km, based on LIGO technology but 
with further improvements.

In this study, we follow the simulation methodology described in 
\cite{fisher3}, adopting the same assumptions and setup for generating the 
GW source catalog. We consider compact binary coalescences including binary 
black holes (BBH), binary neutron stars (BNS) and black hole-neutron star 
binary (BBHNS). The masses of black holes are taken to be uniform in the 
interval $[10,30]M_\odot$, whereas the masses of neutron stars lie in the range 
$[1,2]M_\odot$, under the conditions $m_1 \gtrsim m_2$ and $\eta < 0.25$. 
Also, the spin magnitudes of BH are considered to be uniformly distributed in 
the range $[-1,1]$, while neutron stars are assumed to be non-spinning 
$\chi_{\text{NS}}=0$. In this analysis we consider $\Omega_{m0}=0.308$ and 
$H_0=67.8\, \text{kms}^{-1}\text{Mpc}^{-1}$ \cite{planck_value}. The redshift 
distribution function is considered to be of the form \cite{fisher3,redshift}:
\begin{equation}
P(z) \propto \frac{4\pi d^{\,2}_c(z)\,r(z)}{H(z)(1+z)},
\end{equation}
where $d_c(z)\equiv \int_{0}^{z} \frac{dz'}{H(z')}$ is the comoving distance 
between the source and detector, and $r(z)$ describes the time evolution of 
the burst rate, defined as
\begin{equation}
r(z) =
\begin{cases}
1 + 2z, & z \leq 1, \\
\frac{3}{4}(5 - z), & 1 < z \leq 5, \\
0, & z > 5.
\end{cases}
\end{equation}
Using the above redshift distribution function, we generate source redshifts, 
and for each simulated redshift, we sample GW events with the specified mass 
and spin distributions. For each simulated event, we construct the GW signal 
using the TaylorF2 waveform model \cite{PN_waveform}. The SNR is then computed for each 
event using the detector noise power spectral density, and only events with 
$\rho>8$ are considered as the detected GW events. Signals below this 
threshold are difficult to distinguish from detector noise and may not 
satisfy the high SNR approximation of fisher analysis approach. We then 
calculate the Fisher matrix to estimate the uncertainties in the model 
parameter. Following this methodology, we present and discuss the statistical 
results in the following section.
\section{Results and discussions}
\label{sec6}
In this section, we present the results of our Fisher analysis for compact 
binary coalescences using third generation ground-based GW detectors. We 
analyse the detection capabilities and parameter estimation accuracy of both 
the detectors, ET and CE, respectively.
\subsection{Signal-to-noise ratio and detection}
\begin{figure}[!h]
    \centerline{
    \includegraphics[scale = 0.4]{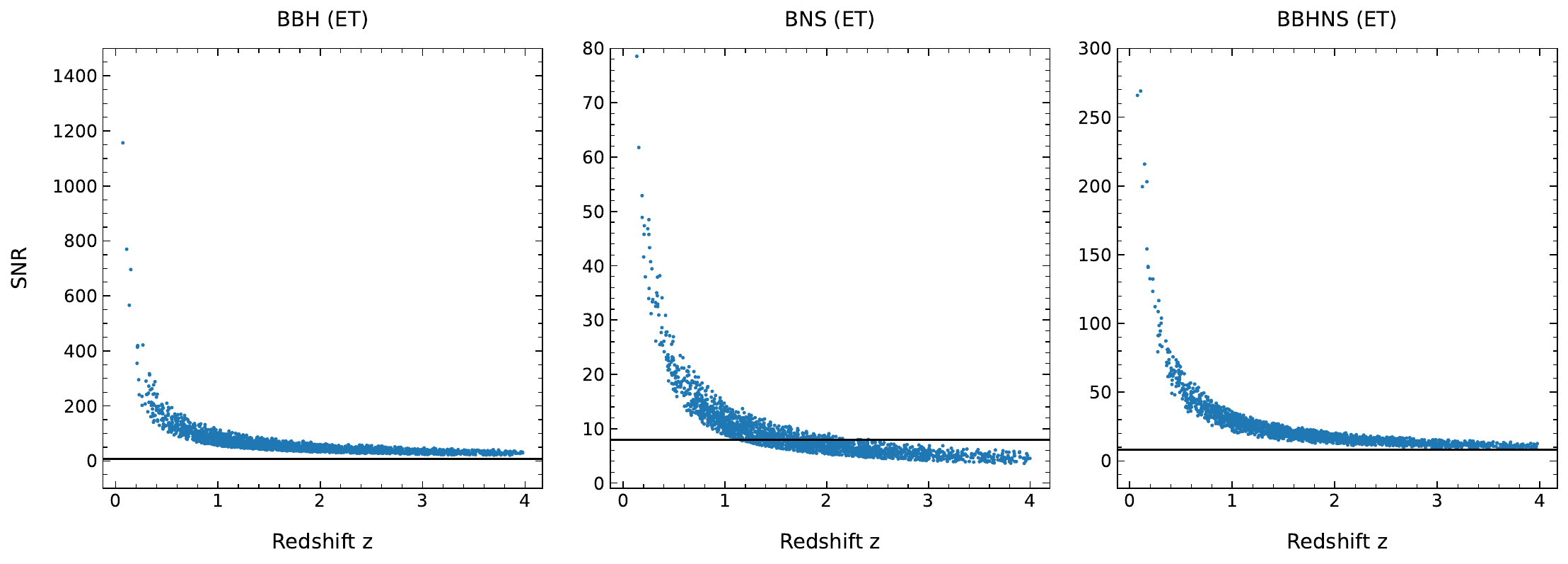}}
    \vspace{-0.3cm}
    \caption{Variation of signal-to-noise ratio (SNR) with redshift for 
several simulated BBH (left panel), BNS (middle panel) and BBHNS (right panel) 
events as observed by Einstein Telescope (ET). The horizontal black line in 
each panel represents the detection threshold SNR $=8$.}
    \label{fig3}
\end{figure}

\begin{figure}[!h]
    \centerline{
    \includegraphics[scale = 0.4]{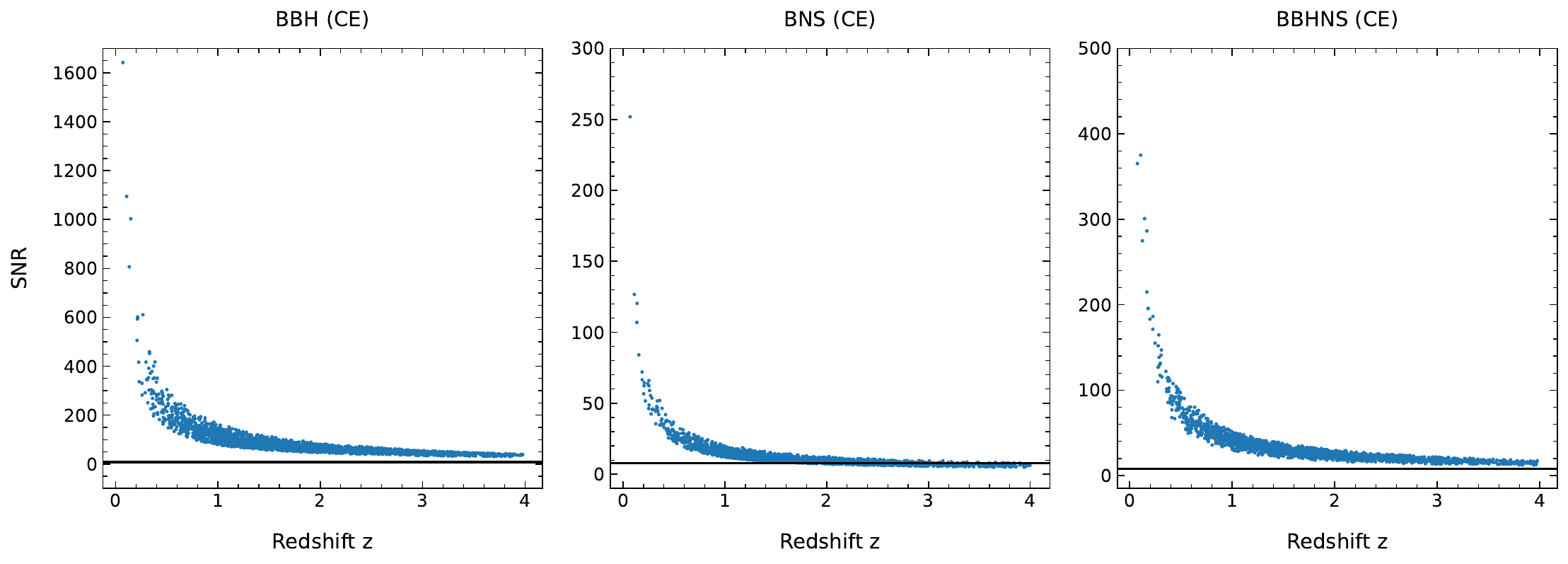}}
    \vspace{-0.3cm}
    \caption{Variation of signal-to-noise ratio (SNR) with redshift for 
several simulated BBH (left panel), BNS (middle panel) and BBHNS (right panel) 
events as observed by Cosmic Explorer (CE). The horizontal black line in each 
panel represents the detection threshold SNR $=8$.}
    \label{fig4}
\end{figure}

We examine the SNR for different compact binary coalescences as a function of 
redshift $z$. For ET, which is composed of three independent interferometers, 
the combined SNR for the network of these independent interferometers is given 
by \cite{redshift}
\begin{equation}
\rho = \sqrt{ \sum_{j\,=\,1}^{3} \left[ \rho^{(j)} \right]^2}.
\end{equation}
In contrast, the CE is designed as a single L-shaped interferometer, and its 
SNR is computed using the standard single detector expression. Figs.~\ref{fig3}
and \ref{fig4} present the SNRs of BBH, BNS and BBHNS as a function of 
redshift for ET and CE, respectively, with the analysis restricted to 
$0 \leq z \leq 4$. From these results (as shown in Figs.~\ref{fig3}
and \ref{fig4}), it is evident 
that BBH and BBHNS events can be detected up to the highest redshift value 
considered with appreciable SNR in both detectors. In case of BNS, ET is 
sensitive up to $z \sim 1.5$, beyond which the SNR drops below the typical 
detection threshold ($\rho=8$), whereas CE retains its sensitivity to 
significantly high redshift values. Moreover, CE consistently yields higher 
SNR values compared to ET across all source classes, implying greater 
precision of parameter estimation.
\subsection{Parameter estimation for Einstein Telescope}
Following the SNR analysis presented above, we now investigate the parameter 
estimation capability of ET. We simulated about $\mathcal{O}(10^{4})$ compact 
binary coalescence events for each class of sources. This choice is motivated 
by the expected detection rates of third generation GW detectors, which are 
anticipated to observe a large number of compact binary coalescences over 
their operational lifetime \cite{ET5, CE2}. Using definition \eqref{eq21}, we 
calculate the Fisher matrix for each class of GW events distributed across 
the redshift range considered. From the Fisher matrices, we can calculate the 
1-$\sigma$ uncertainties on the parameters using equation \eqref{eq20}. For 
BBH systems, we find that ET can constrain the model parameter ${\beta}/{H_0}$ 
up to $\mathcal{O}(10^{-4})$, whereas for BBHNS and BNS systems, the 
constraints are comparatively weaker, at the level of $\mathcal{O}(10^{-3})$, 
with constraint from BBHNS systems being tighter than that from BNS systems.

\begin{figure}[!h]
    \centerline{
        \includegraphics[scale = 0.39]{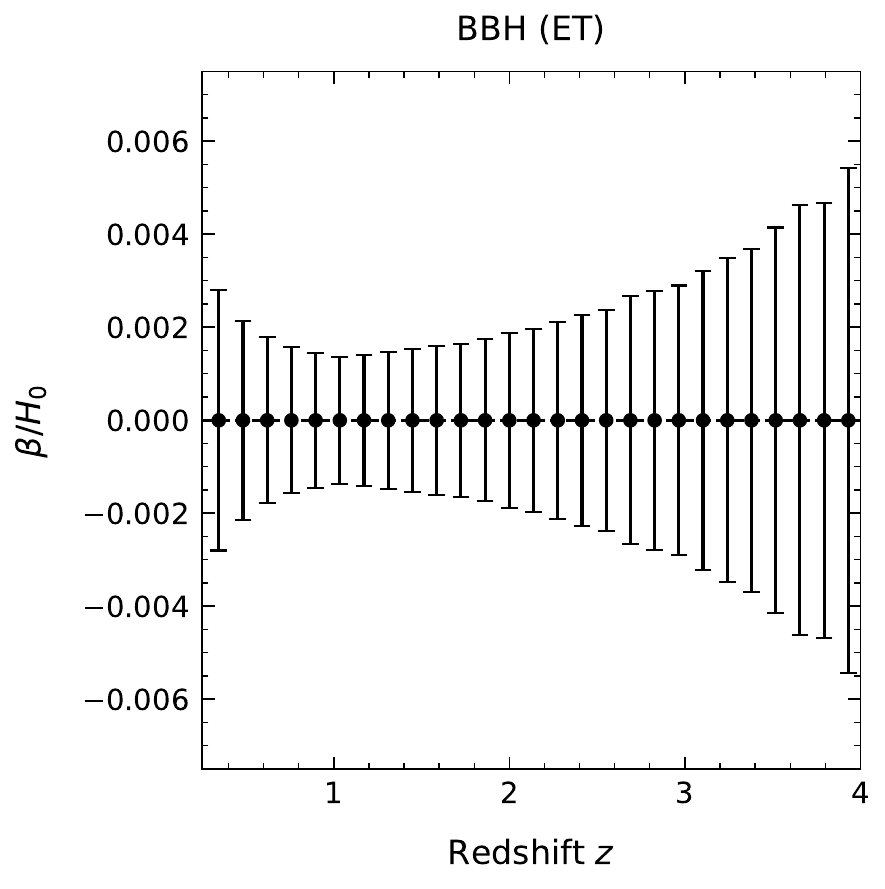}
        \includegraphics[scale = 0.39]{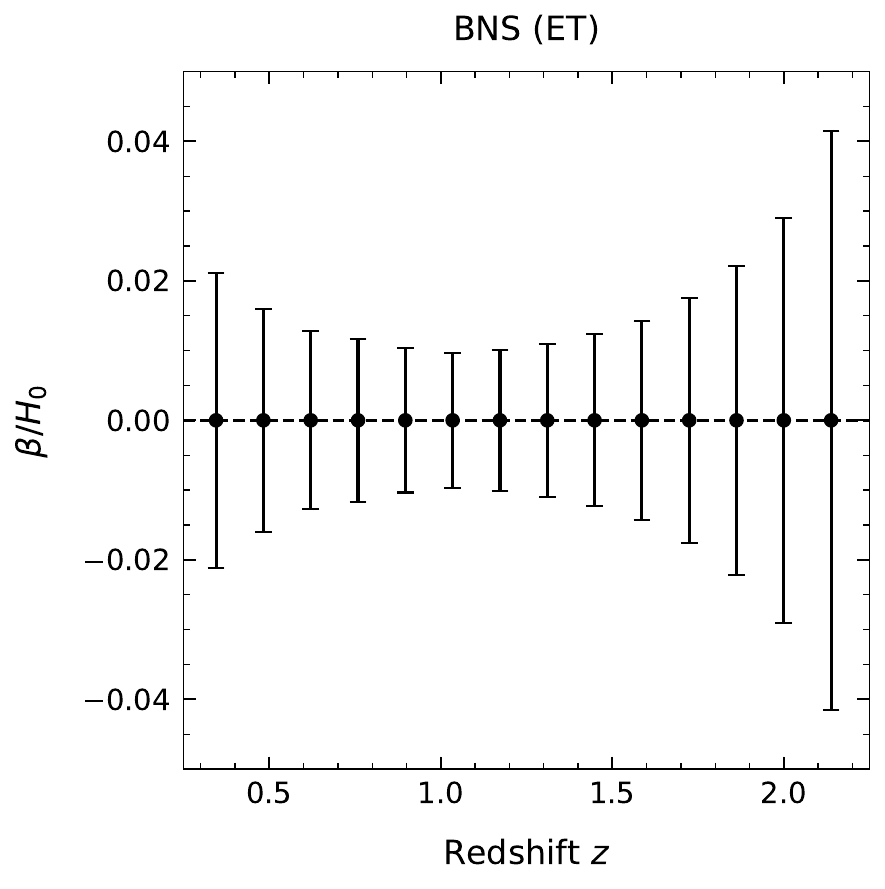}
        \includegraphics[scale = 0.39]{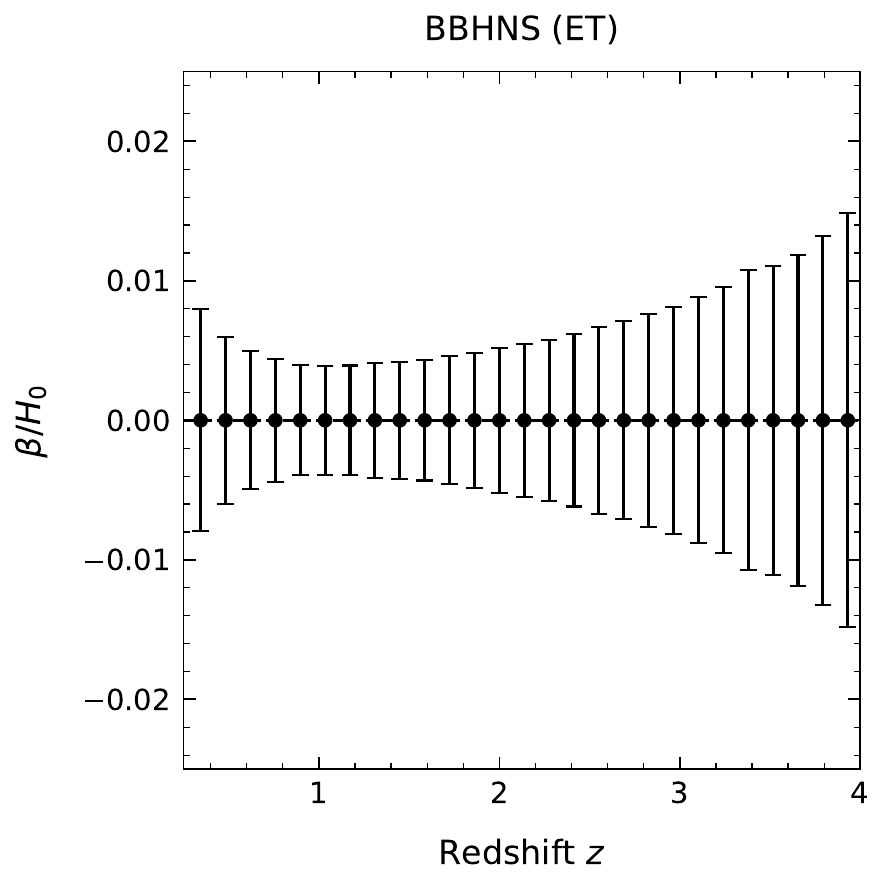}}
        \vspace{-0.2cm}
\caption{Constraints on $\beta/H_0$ obtained by simulating several BBH (left), 
BNS (middle), and BBHNS (right) events in each redshift bin, from the 
perspective of ET power spectral density sensibility.}
    \label{fig5}
\end{figure}

In Fig.~\ref{fig5}, we show the redshift dependence of the 1-$\sigma$ 
uncertainties on the parameter $\beta/H_0$ for the simulated events across 
$z=0$ to $4$. Here, we consider the fiducial value of $\beta/H_0=0$ 
corresponding to GR. For BBH and BBHNS, we observe that the parameter is well 
constrained at the intermediate redshift ($z \sim 0.5-2$), while the 
uncertainties increase towards both the low and high redshifts. The 
upgradation in uncertainties at low redshifts ($z<<1$) is due to the small 
number of detected events in that region. At the high redshift regions, the 
constraints worsen primarily due to a reduction in SNR value, which limits the 
accuracy of parameter estimation. Overall, the constraints for BBHNS are 
weaker in comparison to BBH systems. In contrast, BNS systems have the weakest 
constraint as compared to BBH and BBHNS, which is due to the low SNR value.
\subsection{Parameter estimation for Cosmic Explorer}
We repeat the above analysis, now for CE, considering the CE power spectral 
density sensibility. We find that for BBH and BBHNS systems, CE can constrain 
$\beta/H_0$ at the level of $\mathcal{O}(10^{-4})$, with the magnitude of 
uncertainty for BBHNS greater than that for BBH. On the other hand, for BNS, 
CE can constrain the parameter at the level of $\mathcal{O}(10^{-3})$.

\begin{figure}[!h]
    \centerline{
        \includegraphics[scale = 0.39]{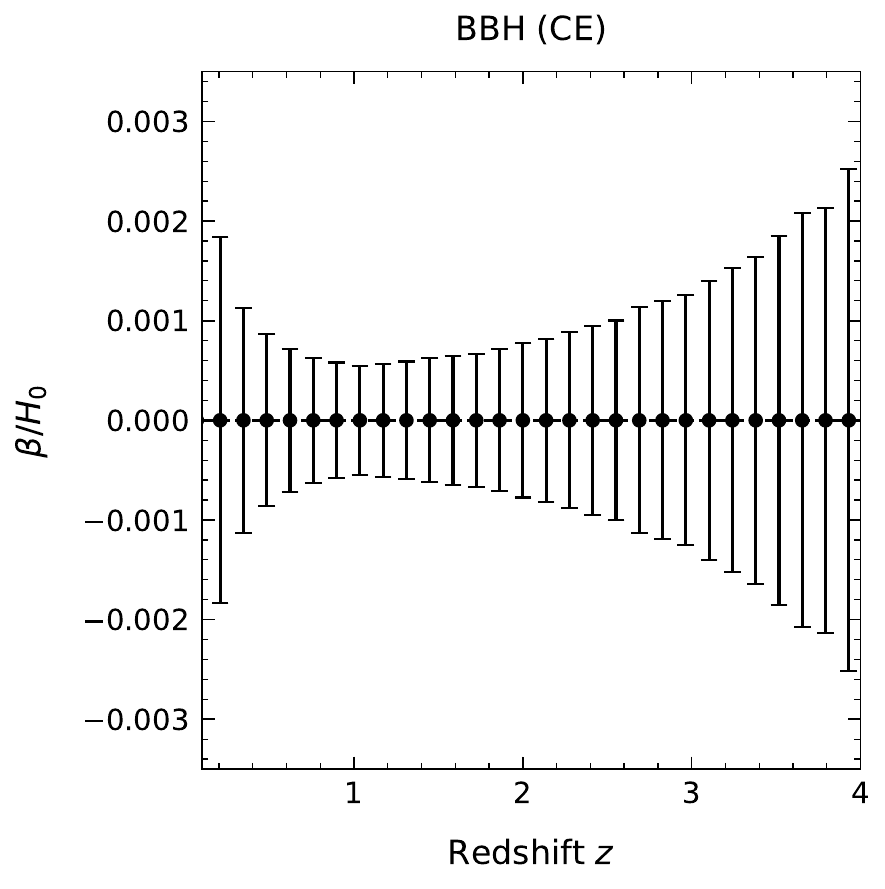}
        \includegraphics[scale = 0.39]{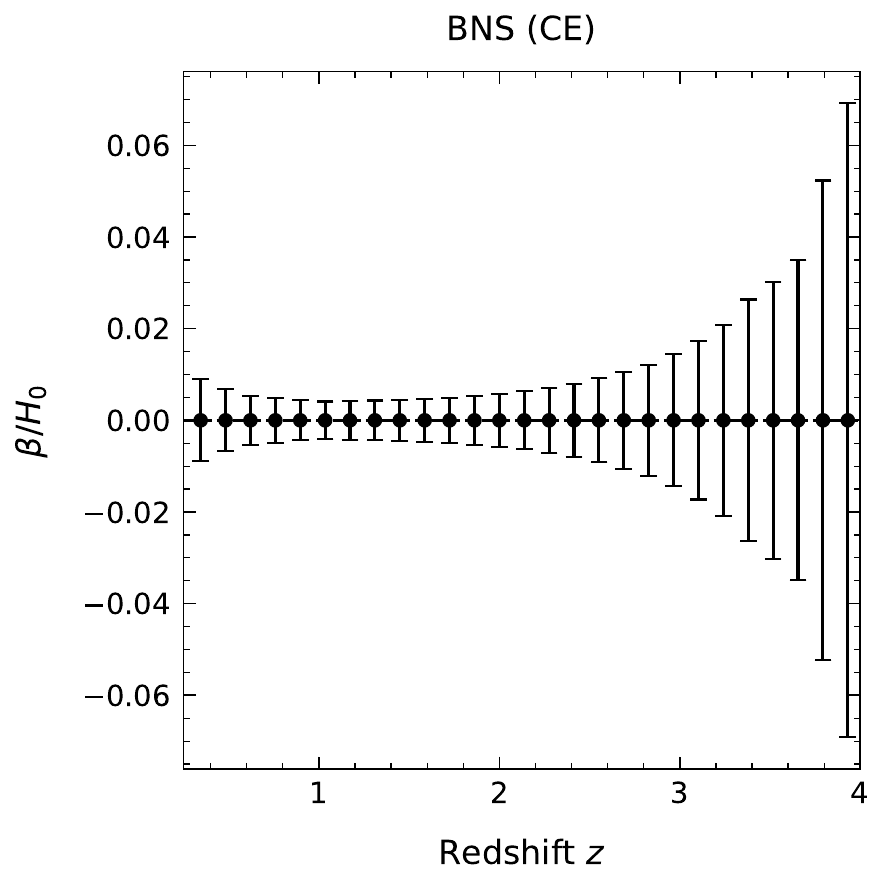}
        \includegraphics[scale = 0.39]{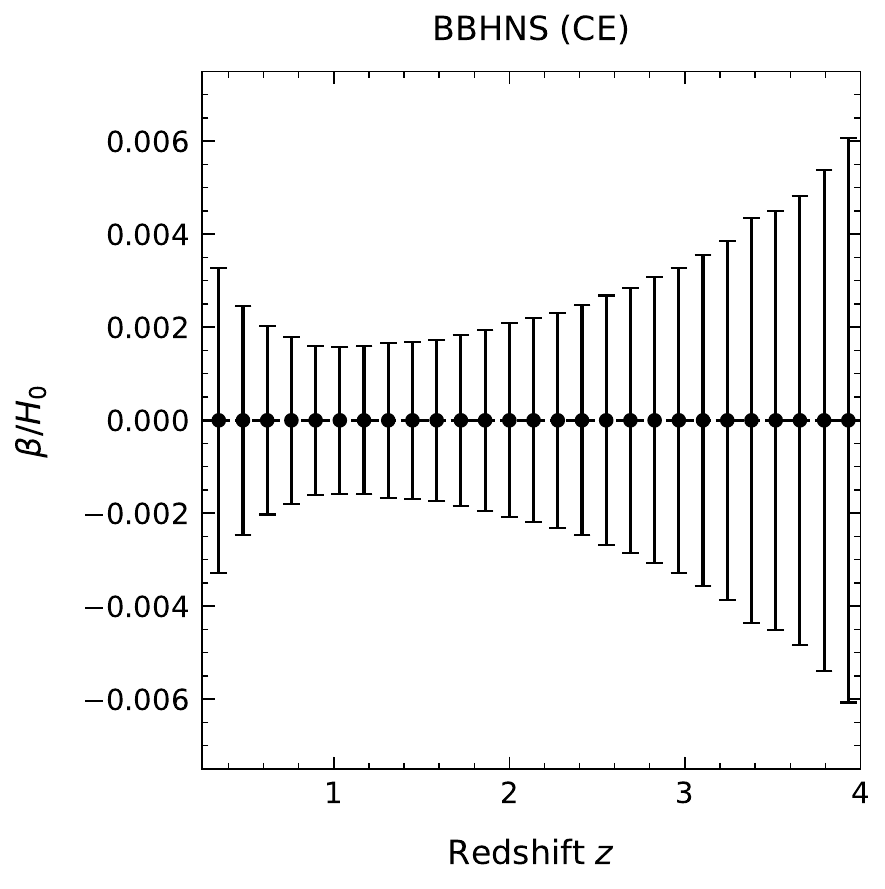}}
        \vspace{-0.2cm}
    \caption{Constraints on $\beta/H_0$ obtained by simulating several BBH 
(left), BNS (middle), and BBHNS (right) events in each redshift bin, from the 
perspective of CE power spectral density sensibility.}
    \label{fig6}
\end{figure}

Fig.~\ref{fig6} shows the redshift dependence of the 1-$\sigma$ uncertainties 
on the parameter $\beta/H_0$ for the simulated binary coalescences, viz., BBH, 
BNS and BBHNS, respectively, across the redshift range $0\leq z\leq 4$, 
considering the fiducial value of $\beta/H_0=0$. As in the case of ET, we 
observe that the constraints are stronger at the intermediate redshifts, and 
the uncertainty increases towards both low and high redshifts. However, 
compared to ET, CE provides better constraints across the entire redshift 
range. This is basically due to its enhanced sensitivity, which leads to 
a higher SNR value and a large number of detectable events across the redshift 
range, particularly in the case of BNS. This enables CE to probe the model 
parameter more effectively in a significantly larger redshift region.
\subsection{Constraints beyond General Relativity}
In the previous subsections, we have constrained the model parameter by taking 
the fiducial value $\beta/H_0=0$. We now consider that the value of 
$\beta/H_0$ lies in the range $[-0.05,0.05]$, which is in agreement with the 
existing constraint on the considered $f(Q)$ gravity model \cite{ss1}. This 
allows us to investigate how accurately the future GW detectors can probe the 
possible deviations from GR, if the underlying gravity theory is indeed 
modified.
 
\begin{figure}[!h]
    \centerline{
        \includegraphics[scale = 0.39]{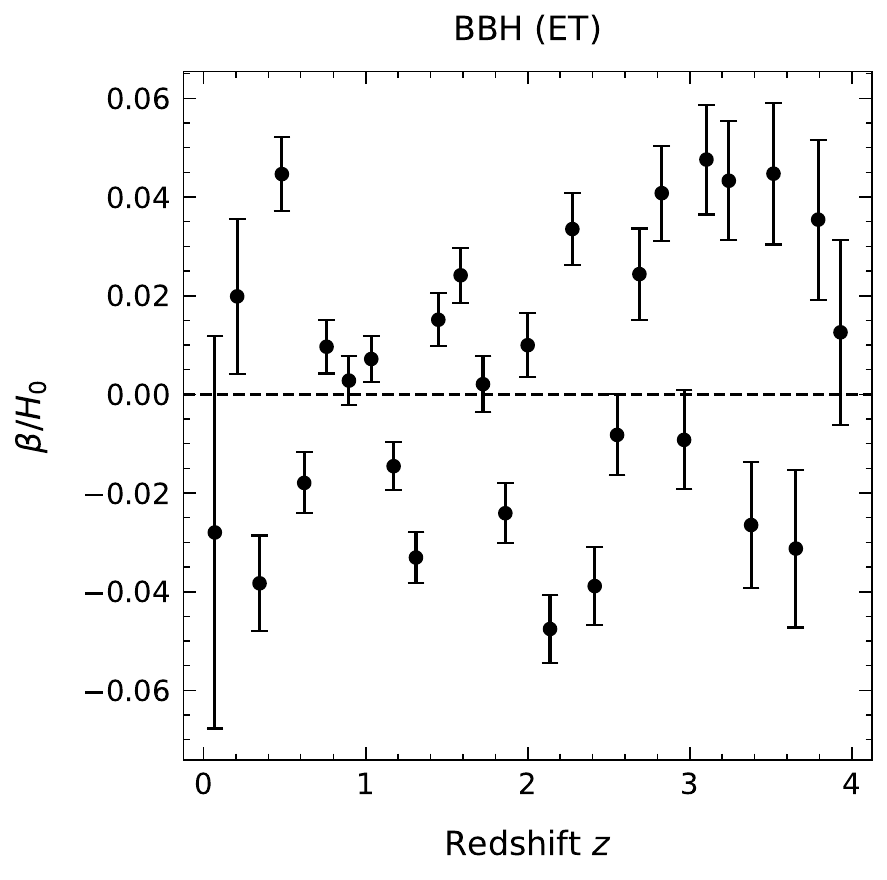}
        \includegraphics[scale = 0.39]{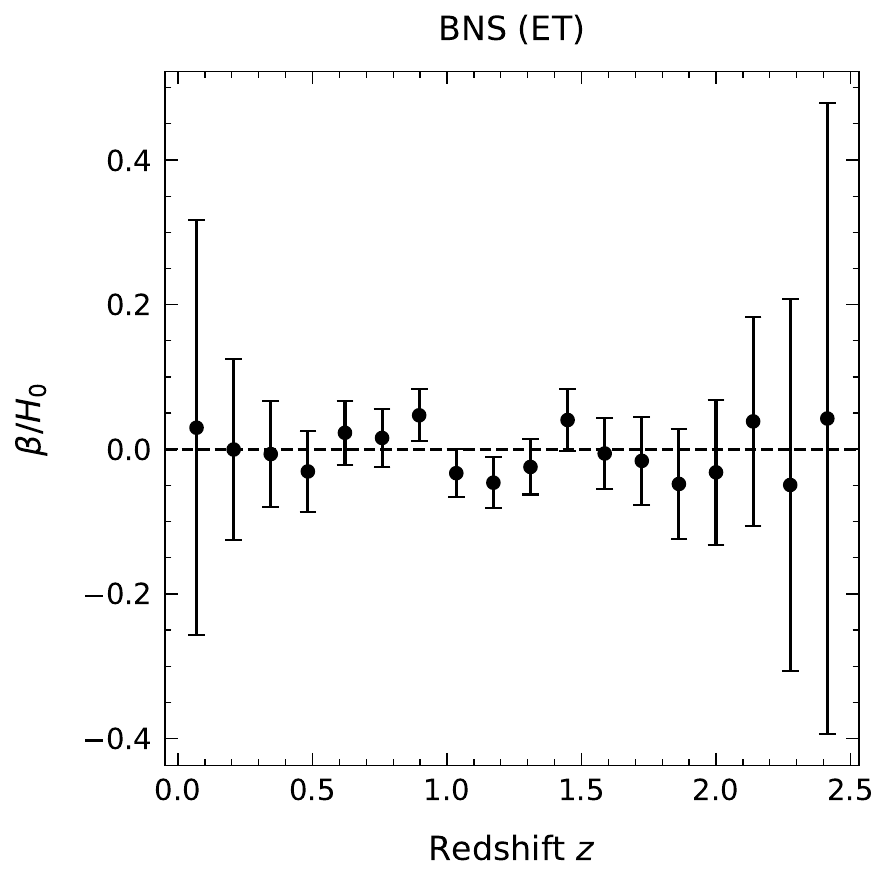}
        \includegraphics[scale = 0.39]{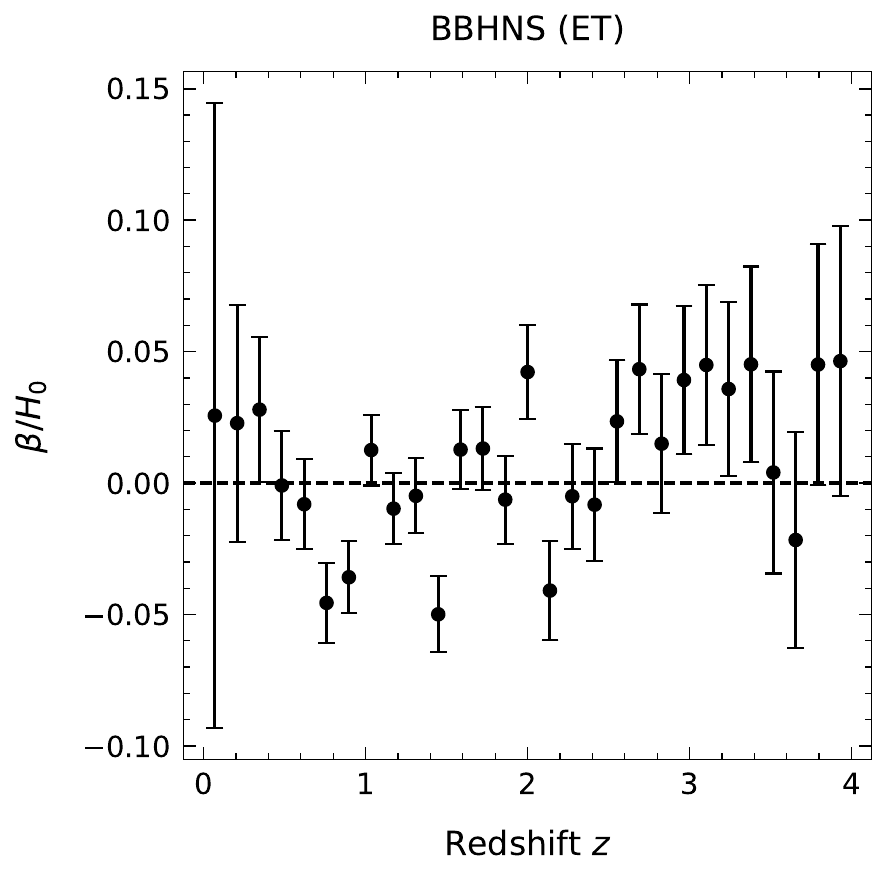}}
        \vspace{-0.2cm}  
    \caption{Constraints on $\beta/H_0$ from a random $\beta/H_0$ sample 
obtained by simulating several BBH (left), BNS (middle), and BBHNS (right) 
events in each redshift bin, from the perspective of ET power spectral density 
sensibility.}
   \label{fig7}
\end{figure}

\begin{figure}[!h]
    \centerline{
        \includegraphics[scale = 0.39]{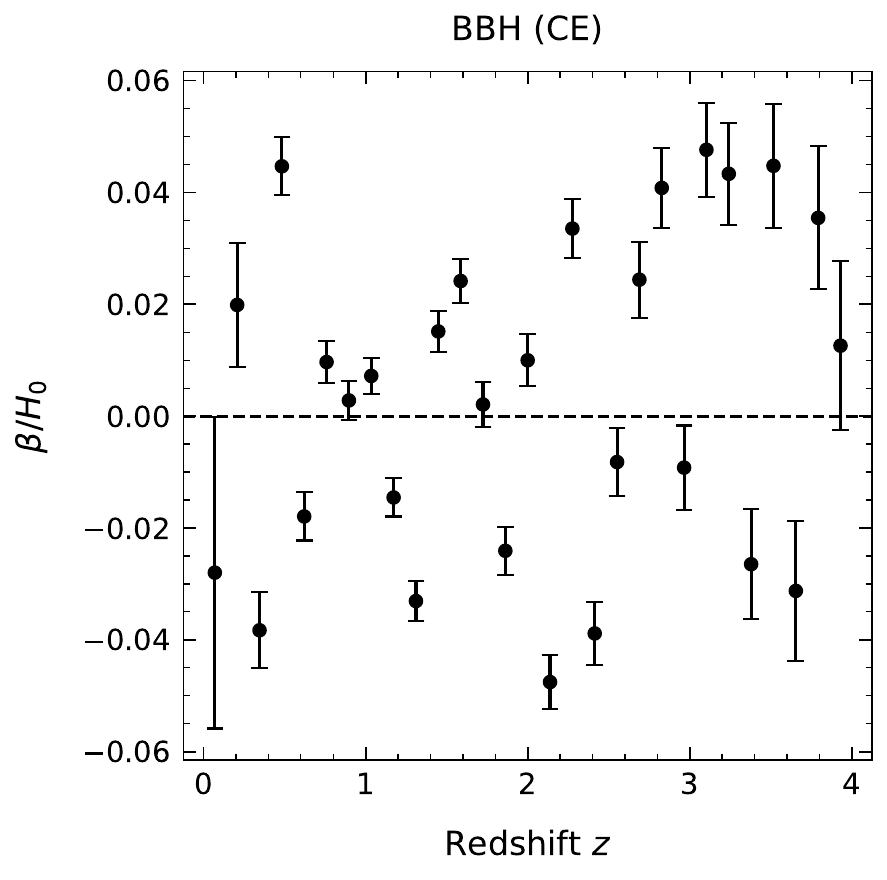}
        \includegraphics[scale = 0.385]{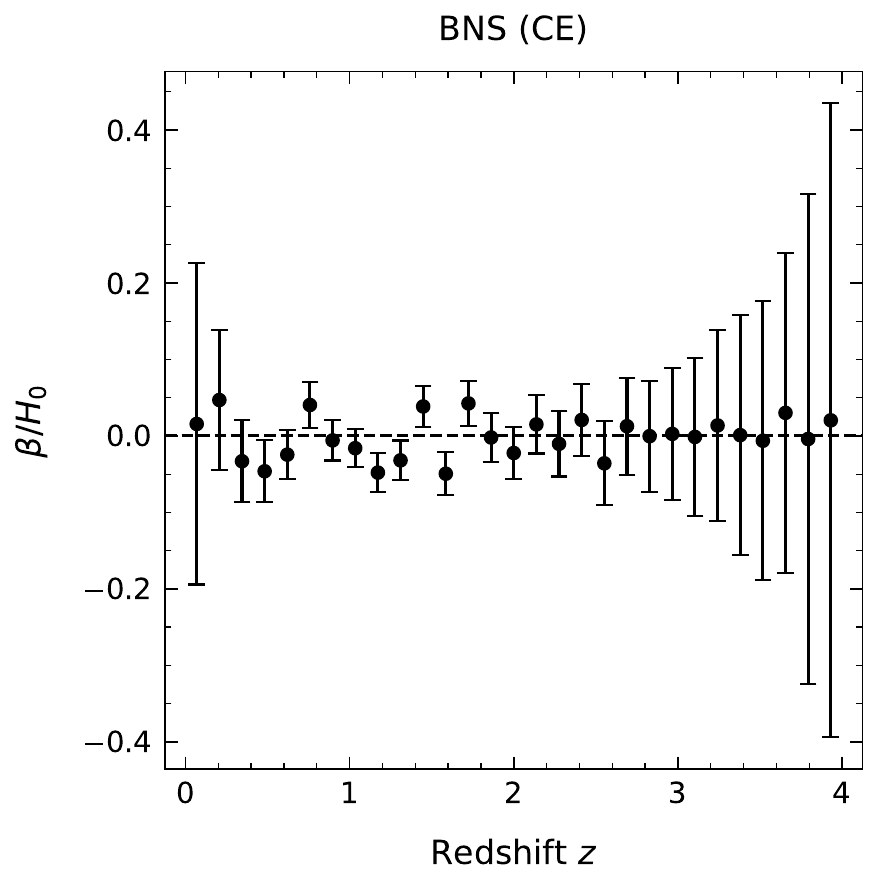}
        \includegraphics[scale = 0.40]{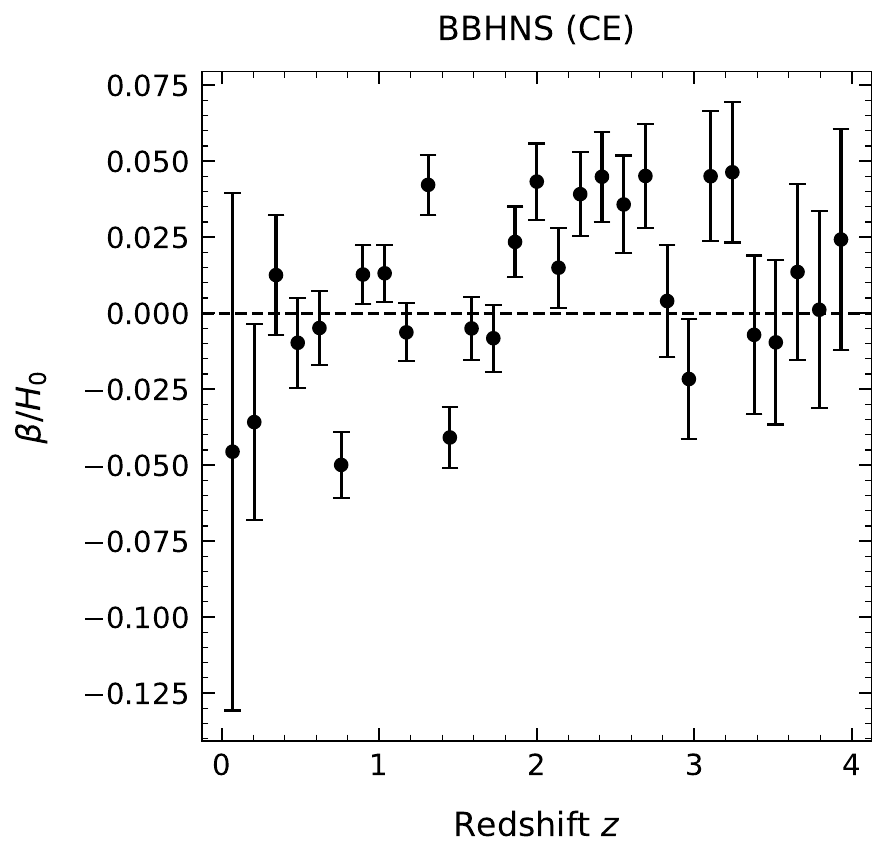}}
        \vspace{-0.2cm}
    \caption{Constraints on $\beta/H_0$ from a random $\beta/H_0$ sample 
obtained by simulating several BBH (left), BNS (middle), and BBHNS (right) 
events in each redshift bin, from the perspective of CE power spectral density 
sensibility.}
   \label{fig8}
\end{figure}

Fig.~\ref{fig7} shows the 1-$\sigma$ uncertainty on the parameter from the 
random sample of $\beta/H_0$ assuming several GW events in each redshift bin 
within ET power spectral noise. From the figure, it is evident that ET 
observations of BBH events possess the capability to distinguish a nonzero 
$\beta/H_0$ from GR, however, BBHNS and BNS events are well compatible with 
$\beta/H_0 \simeq 0$. Similarly, in the case of CE (Fig.~\ref{fig8}), 
BBH events provide a strong constraint on the non-zero values of $\beta/H_0$, 
whereas the constraining ability progressively weakens for BBHNS and BNS 
events.  
\section{Summery and Conclusion}
\label{sec7}
In this work, we have forecast constraints on the model parameter of a $f(Q)$ 
gravity model that mimics a $\Lambda$-CDM background, using modified 
gravitational waveforms from compact binary coalescences, viz., BBH, BNS and 
BBHNS systems. We considered two next-generation ground-based detectors, ET 
and CE, respectively, whose sensitivities are expected to be significantly 
better than those of current GW detectors, and employed the Fisher analysis 
technique to estimate the respective bounds for each binary coalescence.

Our results show that the future ground-based detection of GWs from different 
binary coalescences can place tighter constraints on the model parameter and 
thereby provide a powerful probe to search for the possible deviation of the 
model from GR. In particular, the constraints obtained from inspiraling compact 
binary waveforms are found to be significantly stronger than those obtained in 
previous studies based on standard siren, the luminosity distance 
measurements \cite{ss1}. However, a direct comparison between the two 
results should be treated with caution, as there is a statistical difference 
between the two studies. Nevertheless, the strong constraint is obtained 
because, in the waveform-based approach, 
the parameter is constrained directly from the full gravitational waveform, 
which incorporates both amplitude and phase information. In contrast, standard 
siren analyses rely only on luminosity distance measurements, leading to 
larger degeneracies with cosmological parameters and consequently weaker 
constraints. Overall, our results show the potential of the next-generation 
GW detectors to test ATGs with high precision.

Our work is restricted to the TaylorF2 PN inspiral 
gravitational waveform only for an $f(Q)$ gravity model. However, a natural 
extension of this work would be to compute bounds on the model parameters 
using ringdown phase GW signals as well as the complete 
inspiral-merger-ringdown waveform. Furthermore, extending this study to space 
based detector LISA could provide constraints to a wider frequency range and 
enhance our ability to probe deviations from GR. Moreover, such studies can 
be extended to models of different gravity frameworks beyond GR.

\end{document}